\documentclass[doublecol]{epl2} 

\usepackage{graphicx}
\usepackage{amsmath, amssymb}
\usepackage{lmodern}
\usepackage[T1]{fontenc}
\usepackage[utf8]{inputenc}
\usepackage[english]{babel}
\usepackage{float}
\usepackage{color}

\title{Spatial patterns and biodiversity in off-lattice simulations of a cyclic three-species Lotka-Volterra model}

\author{P.P. Avelino\inst{1,2} \and D. Bazeia\inst{3} \and L. Losano\inst{3} \and J. Menezes\inst{1,4,5} \and B.F. de Oliveira\inst{6}}

\institute{                    
  \inst{1} Instituto de Astrof\'{\i}sica e Ci\^encias do Espa{\c c}o, Universidade do Porto, CAUP, Rua das Estrelas, PT4150-762 Porto, Portugal \\
  \inst{2} Departamento de F\'{\i}sica e Astronomia, Faculdade de Ci\^encias, Universidade do Porto, Rua do Campo Alegre 687, PT4169-007 Porto, Portugal \\
  \inst{3} Departamento de F\'{\i}sica, Universidade Federal da
Para\'{\i}ba 58051-900 Jo\~ao Pessoa, PB, Brazil \\
  \inst{4} Escola de Ci\^encias e Tecnologia, Universidade Federal do Rio Grande do Norte\\
Caixa Postal 1524, 59072-970, Natal, RN, Brazil \\
  \inst{5} Institute for Biodiversity and Ecosystem
Dynamics, University of Amsterdam, Science Park 904, 1098 XH
Amsterdam, The Netherlands \\
  \inst{6} Departamento de F\'{\i}sica, Universidade Estadual de
Maring\'a, Av. Colombo 5790, 87020-900 Maring\'a, PR, Brazil 
}

\abstract{
Stochastic simulations of cyclic three-species spatial predator-prey models are usually performed in square lattices with nearest neighbor interactions starting from random initial conditions. In this Letter we describe the results of off-lattice Lotka-Volterra stochastic simulations, showing that the emergence of spiral patterns does occur for sufficiently high values of the (conserved) total density of individuals. We also investigate the dynamics in our simulations, finding an empirical relation characterizing the dependence of the characteristic peak frequency and amplitude on the total density. Finally, we study the impact of the total density on the extinction probability, showing how a low population density may jeopardize biodiversity.}

\begin{document}

\maketitle

Cyclic predator-prey models, so-called rock-paper-scissors (RPS) models, have provided insight into some of the crucial mechanisms responsible for biodiversity \cite{2002-Kerr-N-418-171,2004Natur.428..412K,2006-Reichenbach-PRE-74-051907-a,2007PhR...446...97S,2007-Reichenbach-N-488-1046,2007-Reichenbach-PRL-99-238105-a,2008-Reichenbach-JTB-254-368-a,2010-Frey-PA-389-4265,2010-He-PRE-82-051909,2011-He-EPJB-82-97,2011-Jiang-PRE-84-021912,2012-Dobrinevski-PRE-85-051903-a,2013-Knebel-PRL-110-168106,2013-Vukov-PRE-88-022123-a,2016-Szolnoki-SR-6-38608} (see also \cite{1920PNAS....6..410L,1926Natur.118..558V,May-Leonard} for the pioneer work by Lotka and Volterra, and May and Leonard).  In their simplest version, spatial RPS models describe the space-time evolution of populations of three different species subject to nearest-neighbor cyclic predator-prey interactions. Simulations of spatial RPS models are usually performed on a square lattice (see, however, \cite{2004-Szolnoki-PRE-70-037102-a} for other lattice configurations). In three-state versions of these models, each site is occupied by a single individual of one of the three species, and there is a conservation law for the total number of individuals, or equivalently, for the total density (these models, involving simultaneous predation and reproduction, are known as Lotka-Volterra models \cite{1920PNAS....6..410L,1926Natur.118..558V}). In four-state versions each site may either be occupied by a single individual or an empty space, and the total density is, in general, no longer conserved (see \cite{2007-Washenberger-JPCM-19-065139} for a case in which the number of individuals per site can be larger than unity and \cite{2008-Szabo-PRE-77-041919-a,2011-Allesina-PNAS-108-5638-a,2011-Durney-PRE-83-051108-a,2012-Avelino-PRE-86-031119,2012-Avelino-PRE-86-036112,2013-Roman-PRE-87-032148,2014-Avelino-PLA-378-393,2014-Avelino-PRE-89-042710,2014-Szolnoki-JRSI-11-0735-a,2017-Avelino-PLA-381-1014} for RPS generalizations involving an arbitrary number of species).

For small enough mobility rates, both three and four-state versions of spatial RPS models have been shown to lead to the stable coexistence of all three species. However, the complex spiralling patterns, observed in stochastic simulations of four-state spatial RPS models, appear to be absent in square lattice simulations of the three-state version.  Furthermore, it has been claimed in \cite{2008-Peltomaki-PRE-78-031906} (see also \cite{2008-Reichenbach-PRL-101-058102-a,2013-Szczesny-EPL-102-28012,2017-Brown-PRE-96-012147}) that the emergence of stable spiral patterns cannot happen in the presence of a conservation law for the total density of individuals. However, as we shall confirm in this letter, in off-lattice simulations, in which the individuals are free to move in a continuous spatial area, the emergence of spiral patterns may occur even in the presence of conservation law for the total density of individuals \cite{2010-Ni-C-20-045116, 2010-Ni-PRE-82-066211} (see also \cite{katerina, PhysRevLett.105.248105,PhysRevB.77.134206} for various applications of off-lattice simulations).

In this letter we describe the results of off-lattice stochastic simulations of spatial RPS models, by confirming that spiral patterns may form in the case of a large enough constant total density. We shall also investigate the dynamical properties of our simulations, quantifying the dependence of the characteristic peak frequency and amplitude on the total density, and study the impact that the total density may have on the conservation of biodiversity.

\section{Off-lattice RPS simulations}

In our off-lattice stochastic simulations $N_A$, $N_B$ and $N_C$ individuals of the species $A$, $B$ and $C$, respectively, are initially randomly distributed on a square-shaped box of linear size $L=1$ with periodic boundary conditions ($N_A = N_B = N_C= N/3$ at the initial time). At each time step a randomly picked individual $I$ of an arbitrary species $S$ moves or preys with probability $m$ or $p=1-m$, respectively. For the sake of definiteness, in this letter, we shall assume that both actions have the same probability of being selected ($m=p=1/2$). Nevertheless, we have verified that this particular choice does not have a significant impact on our main results.

Whenever mobility is selected, a random direction is chosen and the individual $I$ moves in this direction by a distance $\ell_m$. On the other hand, if predation is selected then the individual $I$ looks for the closest prey inside a circular area of radius $\ell_p$ around itself and replaces it by an individual of its own species $S$. If no prey is found within this radius, then the action is not executed. In this letter, we shall make the reasonable assumption that the mobility and predation length scales are identical ($\ell=\ell_m=\ell_p$) and choose $\ell=2 \times 10^{-2}$ (we will later show that our main results are not affected by this specific choice). Note that, unlike in standard RPS lattice simulations, in our off-lattice simulations position swaps between neighbours never occur. 

A generation timescale $\Delta t=1$ is defined as the time necessary for $N$ actions to be realized. In this letter we consider simulations with different values of the (constant) total density $\varrho= N/L^2$ in the interval $[2.4 \times 10^1,3 \times 10^5]$. All simulations have a total duration of $t=1.5 \times 10^4$ generations.

\begin{figure}
\centering
\includegraphics{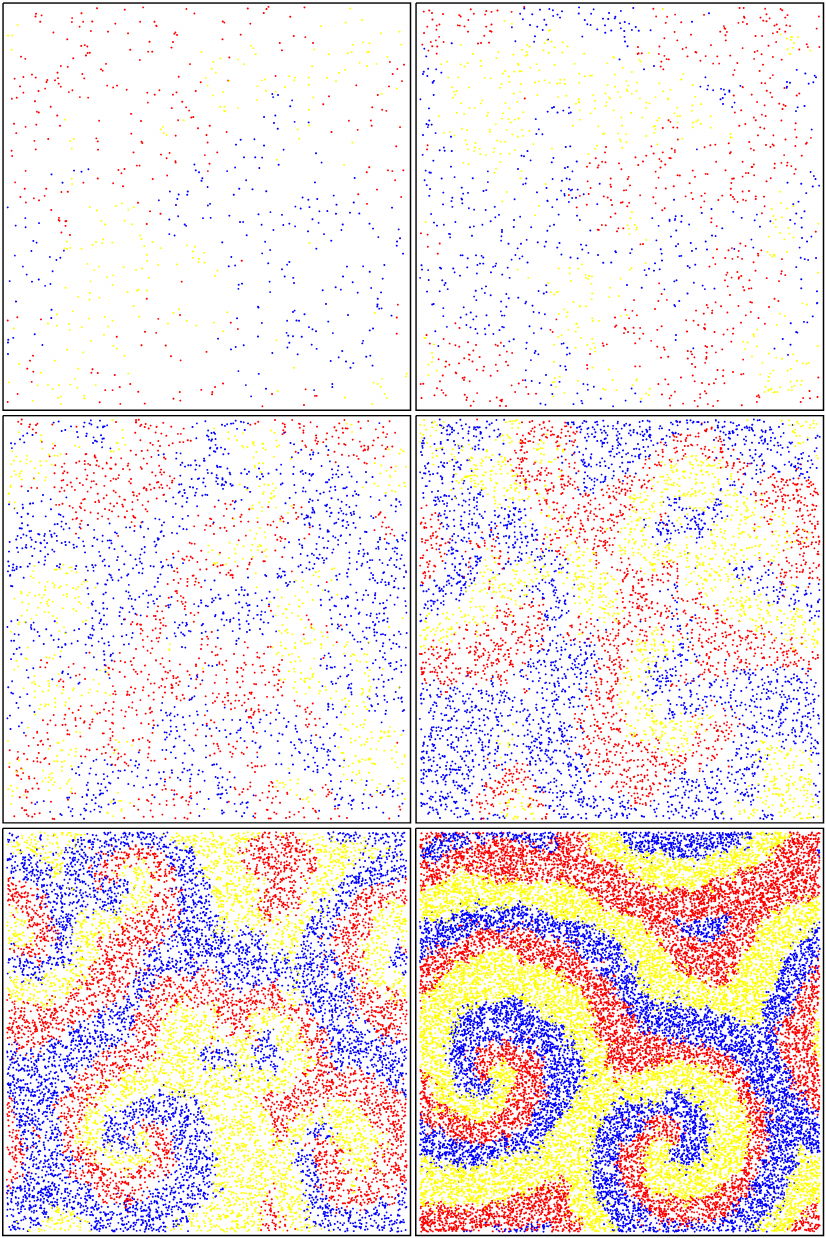}
\caption{Snapshots of our off-lattice stochastic RPS simulations after $t=10^4$ generations, for six different values of the \revision{total density} $\varrho$  ($\varrho = 7.0 \times 10^2$, $1.5 \times 10^3$, $3.2 \times 10^3$, $6.7 \times 10^3$, $1.4 \times 10^5$ and $3.0 \times 10^5$, from top to bottom and left to right, respectively)}
	\label{fig1}
\end{figure}

\section{Spiral patterns}
\label{sec3}
\begin{figure}
\centering
\includegraphics{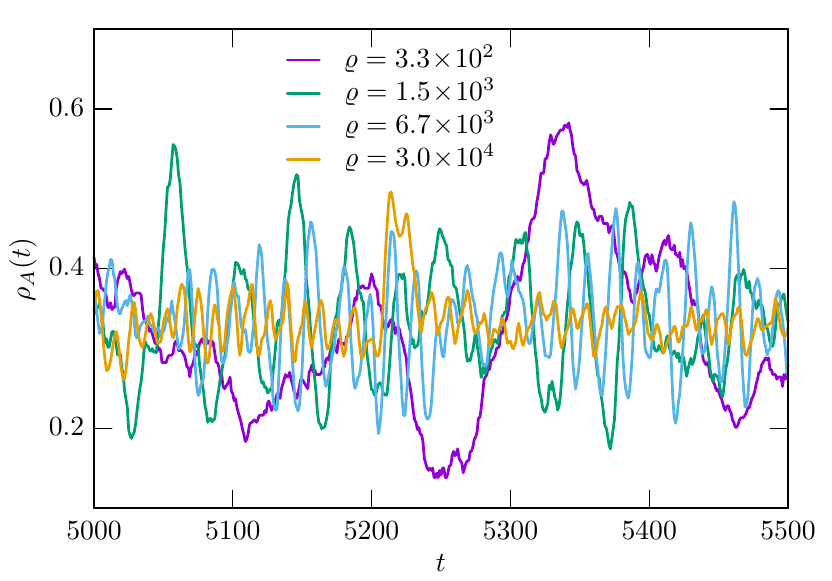}
\caption{Time evolution of the fractional abundance of the species $A$ for different values of the total density $\varrho$ (a similar behavior is found for species $B$ and $C$).}
	\label{fig2}
\end{figure}

Figure~\ref{fig1} shows snapshots taken from six of our off-lattice RPS simulations with constant total densities $\varrho = 7.0 \times 10^2$, $1.5 \times 10^3$, $3.2 \times 10^3$, $6.7 \times 10^3$, $1.4 \times 10^5$ and $3.0 \times 10^5$ (from top to bottom and left to right, respectively) after $t=10^4$ generations. 
It was shown in \cite{2008-Peltomaki-PRE-78-031906}, using lattice-based simulations, that spiral patterns would form only if the total density $\varrho$ is not conserved. However, for the off-lattice simulations described in this letter, spiral patterns may form for sufficiently large values of $\varrho$ even if the total is no longer conserved, as it is shown in Fig.~\ref{fig1}. On the other hand, Fig.~\ref{fig1} shows that spiral formation does not take place for small values of $\varrho$.

If $\ell \ll L$ and $d = \varrho^{-1/2} \ll L$, where $d$ is the characteristic distance between neighbours, the size of the box does not have any significant impact on the patterns which emerge from the simulation, these being dependent essentially on the ratio $q=\ell/d$. Figure~\ref{fig1} shows that spiral patterns are prominent only if $q > 1$. For $q < 1$ the results appear to be consistent with those of the usual stochastic simulations performed in square lattices with nearest neighbour interactions in the presence of a conservation law for the total number of individuals, in which spiral formation is suppressed \cite{2008-Peltomaki-PRE-78-031906}.

\begin{figure}
\centering
\includegraphics{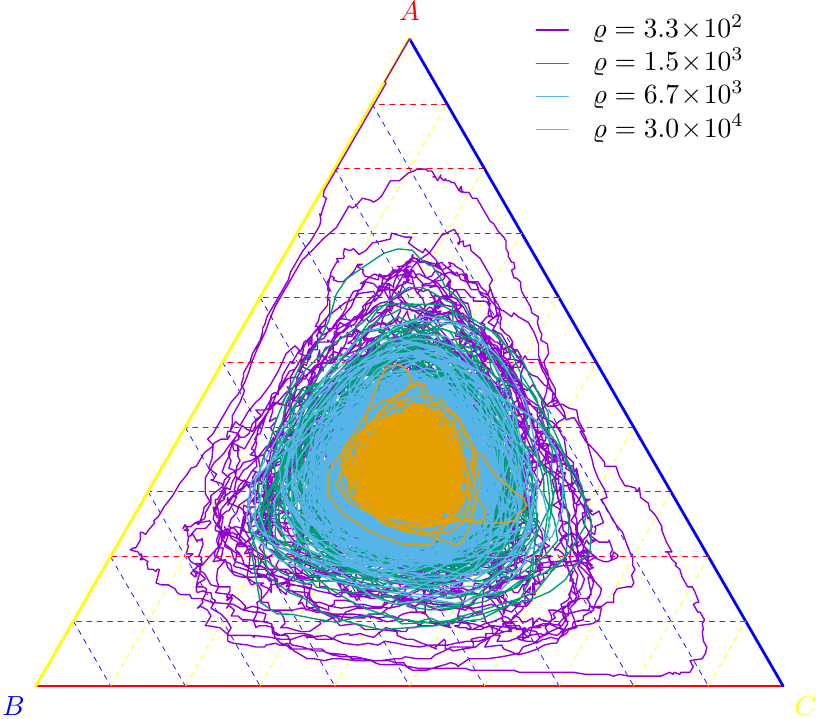}
\caption{The ternary diagram shows the time evolution of the fractional abundances of the three different species on single runs of our simulations for different values of the total density $\varrho$. For $\varrho=3.3 \times 10^2$ only the species $A$ remains by the end of the simulation.}
	\label{fig3}
\end{figure}
\begin{figure}
\centering
\includegraphics{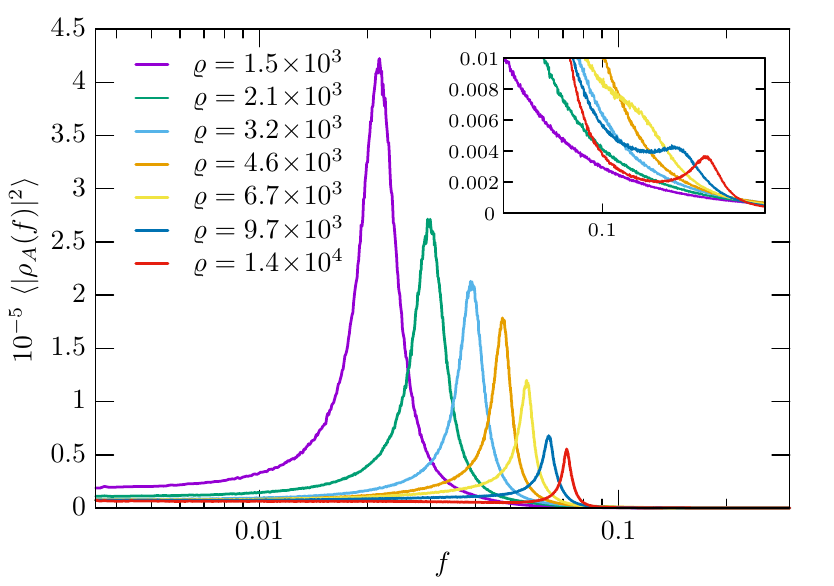}
\caption{Power spectrum of $\rho_A$ for different values of the total density $\varrho$. The results were averaged over $5 \times 10^3$ simulations with a time span $t=1.5 \times 10^4$ generations of different initial conditions (the first $5 \times 10^3$ generations of each simulation have been discarded in the calculation of the power spectrum). Note that a similar behaviour is found for species B and C.}
	\label{fig4}
\end{figure}

\section{Time evolution}

The time evolution of the fractional abundance $\rho_A=N_A/N$ of the species $A$  is shown in Fig.~\ref{fig2} for different values of the total density $\varrho$ (the results for species $B$ and $C$ are analogous --- note that $\rho_A+\rho_B+\rho_C=1$). Figure~\ref{fig2} shows that $\rho_A$ oscillates with a characteristic time and amplitude which depends on the value of the total density $\varrho$. The larger the value of $\varrho$, the smaller the time (measured in units of one generation time) necessary for a predator to find its prey.  In other words,  the larger the value of $\varrho$, the smaller the average distance between a predator and its closest prey. On the other hand, the smaller the value of $\varrho$, the larger the characteristic oscillation amplitude of  $\rho_A$ (see Ref.~ \cite{2009-Tome-CPC-180-536} for a study of the dependence of the oscillation amplitude on $N$ in the context of square-lattice simulations).

The simultaneous representation of the time evolution of the fractional abundances of the three different species $A$, $B$ and $C$ is shown in Fig.~\ref{fig3} for various values of the constant total density $\varrho$. The initial conditions are such that $\rho_A =\rho_B=\rho_C=1/3$, thus implying that all orbits start at the center of the triangle. Fig.~\ref{fig3} shows that the smaller the value of $\varrho$, the larger the area of phase space occupied by the orbits. This implies that the probability of biodiversity being lost increases as $\varrho$ is decreased, thus showing that the impact of the increase of the characteristic oscillation amplitude (as $\varrho$ is decreased) is significantly larger than that of the increase of the characteristic period (or of the consequent reduction of the number of cycles within the simulation time span). In particular, for $\varrho=3.3 \times 10^2$ only the species $A$ remains by the end of the simulation, even though it was on the verge of extinction for several periods before that.  

\begin{figure}[t]
\centering
\includegraphics{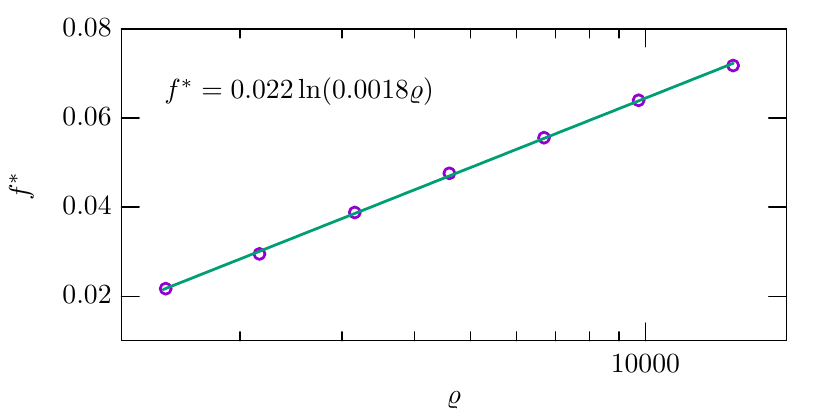}
\includegraphics{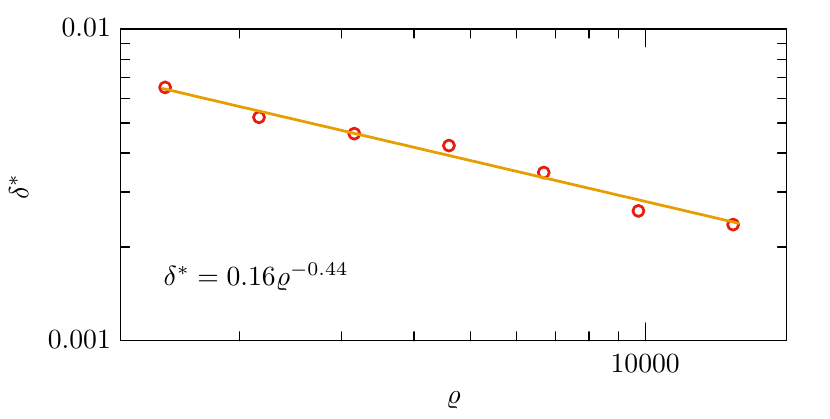}
\caption{Dependence of the peak frequency $f^{*}$ (upper panel) and amplitude $\delta^*$ (lower panel) on the total density $\varrho$. The solid lines represent the fitting functions.}
	\label{fig5}
\end{figure}
\begin{figure}[h]
\centering
\includegraphics{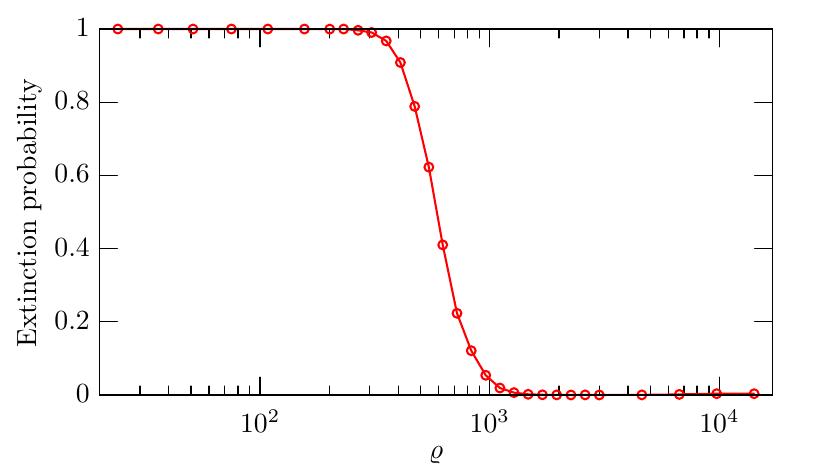}
\caption{The extinction probability as a function of the total density $\varrho$. The results were obtained from $5 \times 10^3$ simulation runs for each data point.}
	\label{fig6}
\end{figure}
In order to provide a more quantitative description of the time evolution of the system, we compute the Fourier transform of the fractional abundance $\rho_A (t)$ of the species $A$. Let us define the temporal discrete Fourier transform as
\begin{equation}
    \rho_A(f) = \displaystyle \frac{1}{N}\sum_{t=t_0}^{N_G} \rho_A(t) \, e^{-2\pi i f
t} \ ,
    \label{dft}
\end{equation}
where $t_0= 5 \times 10^3$, and $N_G = 1.5 \times 10^4$ generations (the first $5 \times 10^3$ generations of each simulation, with the time span $t=1.5 \times 10^4$, have been discarded in the calculation of the power spectrum).

The power spectrum of $\rho_A$ is displayed in Fig.~\ref{fig4} for different values of the total density $\varrho$. The results shown in Fig.~\ref{fig4} were averaged over $5 \times 10^3$ simulations and different initial conditions.  The insert of Fig.~\ref{fig4} also reveals a second peak for high enough values of the total density 
$\varrho$. This is related with the very fast oscillations with small amplitudes also found in 
\cite{2013-Roman-PRE-87-032148}. A similar power spectrum to the one shown in Fig.~\ref{fig4} for species $A$ is also obtained for species $B$ and $C$.

Figure~\ref{fig5} shows the average values, over $5\times 10^3$ simulations, of $f^{*}$ (upper panel) and $\delta^*$ (lower panel) as a function of  the total density $\varrho$. As expected from the previous discussion, the characteristic peak frequency $f^{*}$, defined as the value of the frequency $f$ at the maximum of the power spectrum, increases, while the characteristic peak amplitude, defined by $\delta^*=(\langle|\rho_A(f^*)|^2 \rangle)^{1/2}$, decreases with $\varrho$. Figure~\ref{fig5} also presents an empirical fit quantifying the logarithmic dependence of $f^*$ (upper panel) and the power law dependence $\delta^*$ (lower panel) on the total density $\varrho$.

\section{Average density and biodiversity}

\begin{figure}[h]
\centering
\includegraphics{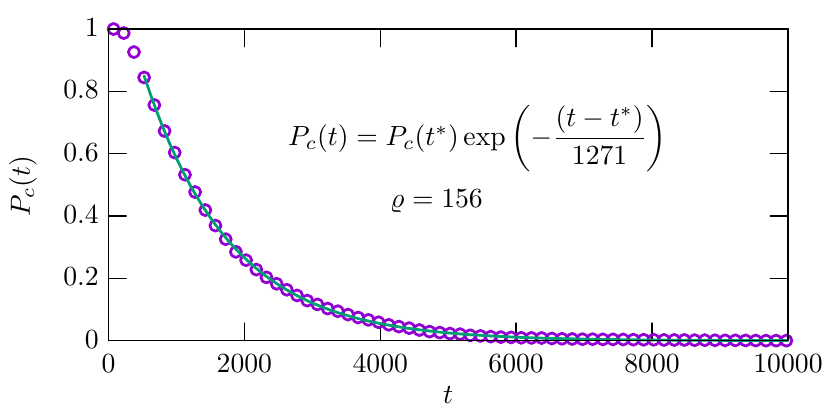}
\caption{The coexistence probability as a function of the time for $\varrho = 156$.}
	\label{fig7}
\end{figure}
The impact of mobility on biodiversity has been investigated in detail using simulations performed in square lattices with nearest neighbour interactions. In these simulations, the ratio $q=\ell/d$ is always equal to unity. We now consider the impact of the total density $\varrho$ on the extinction probability (for a fixed $\ell=0.02$) using our off-lattice simulations, which allow for a variable ratio $q=\ell/d$. To this end, we have performed a set of $5 \times 10^3$ simulations for different values of the total density $\varrho$ and verified whether or not one or more species was extinct after $1.5 \times 10^4$ generations. Figure~\ref{fig6} depicts the extinction probability as function of total density $\varrho$. It shows that biodiversity is maintained only above a critical value of the total density $\varrho_c = (2.3 \pm 0.2) \times 10^2$, with the transition between the coexistence and extinction regimes being quite sharp. 

We also verified that, if $t$ is sufficiently large, the probability $P_c$ of having coexistence after a time $t$ decreases exponentially as
\begin{equation}
P_c(t) = P_c(t^*) e^{-(t-t^*)/{\tau}}\,,
\end{equation}
following a universal law similar to that of radioactive decay. Here $t^*$ is a fixed time and $\tau$ is the average interval of time after which coexistence is lost. In Fig. \ref{fig7} the time evolution of the probability $P_c(t)$ has been estimated from $5 \times 10^3$ simulations with $\varrho=156$. The exponential fit given in Fig.  \ref{fig7}  was obtained after discarding the first three points with $t < 525$ generations.

\section{Conclusions}

In this letter, we reported on the results of off-lattice stochastic simulations of cyclic three-species spatial predator-prey models with a conserved total density. We have shown that spiral patterns do form in these simulations for high values of the total density (that is, for values of the ratio $q=\ell/d$ greater than unity). We have taken advantage of the freedom to vary $q$ in our simulations to study the impact of the total density on population dynamics (for a fixed $\ell$), showing that the characteristic peak frequency and amplitude display, respectively, a logarithmic increase and a power law decrease with the total density. Finally, we have shown that coexistence can only be maintained above a critical value of the total density, with only a narrow transition region between coexistence and extinction regimes, which indicates that even moderate changes on the total density of individuals may have a great impact on the conservation of biodiversity.


\acknowledgments
We thank CAPES, CNPq, Fapern, FCT, Funda\c c\~ao Arauc\'aria, INCT-FCx, and the Netherlands Organisation for Scientific Research (NWO) for financial and computational support. PPA acknowledges support from FCT Grant UID/FIS/04434/2013, DB acknowledges support from Grants CNPq:455931/2014-3 and CNPq:306614/2014-6, LL acknowledges support from Grants CNPq:307111/2013-0 and CNPq:447643/2014-2, and JM acknowledges support from NWO Visitor's Travel Grant 040.11.643.

\end{document}